\documentclass[
  ,final            
  ]
  {aipproc}

\layoutstyle{8x11double}

\begin{document}

\title{PSR J1738+0333: a new gravitational laboratory}


\author{Paulo C. C. Freire}{
  address={N.A.I.C., Arecibo Observatory, HC3 Box 53995, PR 00612,
  U.S.A.; {\tt pfreire@naic.edu}}
}

\author{Bryan A. Jacoby}{
  address={Naval Research Laboratory, Washington, DC; {\tt bryan.jacoby@nrl.navy.mil}}
}

\author{Matthew Bailes}{
  address={Centre for Astrophysics and Supercomputing, Swinburne
  University of Technology, Hawthorn, Australia; {\tt
  mbailes@astro.swin.edu.au}}
}

\begin{abstract}
We describe in this paper a new binary millisecond pulsar,
PSR J1738+0333. Using Arecibo, we have achieved good timing
accuracy for this object, about 220 ns for 1-hour integrations over
100~MHz. This allowed us to measure a precise proper motion, parallax
and orbital parameters for this system. We highlight the system's
potential for constraining alternative theories of gravitation.
\end{abstract}


\maketitle


\section{Introduction}

PSR~J1738+0333 is a 5.85-ms pulsar in a binary system with an orbital
period of 8.5 hours and a companion white dwarf (WD) with a mass of
about 0.2 M$_{\odot}$. This millisecond pulsar (MSP) was found with
the Parkes 64-m Radio Telescope in a 20-cm Multi-Beam search for
pulsars in intermediate Galactic latitudes ( $5^\circ < | b | <
30^\circ$) \cite{jbo+07}; we have been timing it with Arecibo
for the last 4 years using the Wide-band Arecibo Pulsar Processors
(WAPPs, \cite{dsh00}). We have obtained a TOA residual rms of 220
ns per WAPP per hour. This pulsar will be used in the array that is
being used to search for nano-Hertz gravitational waves.

\section{Timing of PSR~J1738+0333}

\begin{figure}
  \label{fig:radial}
  \includegraphics[height=.27\textheight]{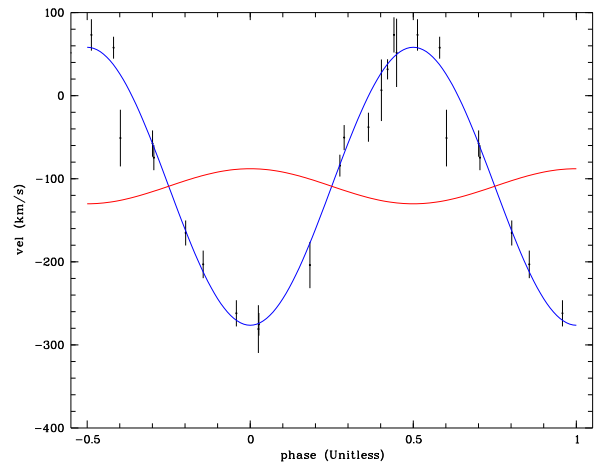}
  \caption{Radial velocity measurements of the companion of
  PSR~J1738+0333 as a function of orbital phase.}
\end{figure}

\begin{figure}
  \label{fig:mass_mass}
  \includegraphics[height=.50\textheight]{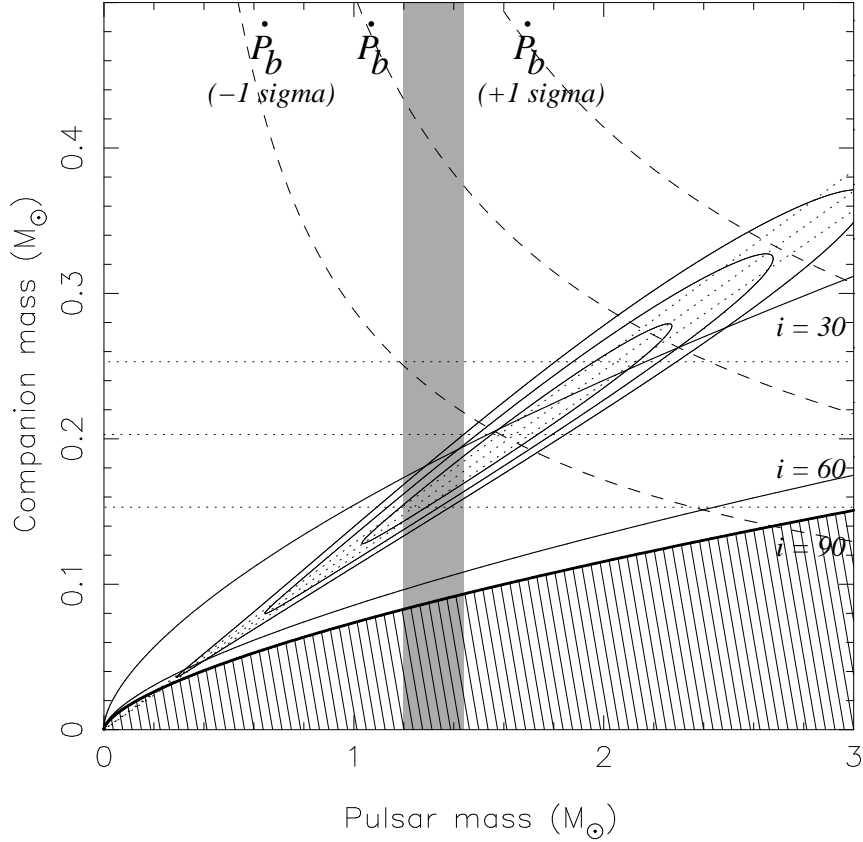}
  \caption{Mass constraints for PSR~J1738+0333 and its white dwarf
  companion. The closed solid lines (contours)
   include 66.3, 95.4 and 99.7\% of the
  probability, this is dictated by the information regarding the
  companion mass (horizontal dotted lines, with $-1,0$ and
  $+1\,\sigma$ limits indicated) and the mass ratio (inclined dotted
  lines, with $-1,0$ and $+1\,\sigma$ limits also indicated. If the
  observed orbital decay, corrected for kinematic effects $\dot{P_b}$
  (dashed lines) is due to emission of quadrupolar gravitational
  waves, then the two regions should overlap, as observed. The
  vertical gray bar indicates the range of precisely measured neutron
  star masses.}
\end{figure}

Initially we sought to determine the companion and
pulsar masses from a measurement of the Shapiro delay. Despite the
high timing precision, the measurement was not possible given the
system's low orbital inclination.

Fortunately, it is possible to determine the masses of the components
independently. This comes from recent optical work of Marten van
Kerkwijk and one of us (BAJ). Using the Magellan telescope on Las
Campanas, Chile, they detected the companion star and found its
spectrum to be similar to that of the companion of PSR~J1909$-$3744;
which has as mass of 0.203 M$_{\odot}$, measured by Shapiro delay
\cite{jhb+05}. For this reason, and from here on, we assume that the
companion of PSR~J1738+0333 has mass ($m_2 = 0.20 \pm 0.05 M_{\odot}$),
but note that a precise estimate of this mass has not yet been made.
This implies an orbital inclination of about $30^\circ$.

Introducing the Shapiro delay that corresponds to this
companion mass and inclination, we obtain an eccentricity of $(8 \pm
16)\times 10^{-8}$, the lowest ever measured for any binary
system. This opens up the possibility of greatly improved tests of the
fundamental nature of spacetime, introducing the most stringent
constraints ever on preferred-frame effects and non-conservation of
momentum \cite{sta03}.

More recently, the radial-velocity curve was measured using Gemini
South on Cerro Pach\'on (see Fig. \ref{fig:radial}).
From this we can derive the mass ratio of the system, $R = 8.1 \pm 0.3$
(Van Kerkwijk, 2007, pers. comm.); therefore the pulsar mass is
$m_1\,=\,(1.7 \pm 0.4) \, \rm M_{\odot}$.
The error estimate for the companion mass assumed above is very
conservative, it admits a wide possible range of companion and pulsar
masses (see Fig. \ref{fig:mass_mass}). It will certainly be
measurable with better precision in the near future, as in the case of
PSR~J1911$-$5958A (\cite{bkkv06,cfpd06}, see also
Bassa et al. these proceedings).

\subsection{A test of the Strong Equivalence Principle}

The measurement of the masses of the components of  important because
it allows a (low-precision) estimate of the expected
rate of orbital decay due to the emission of quadrupolar gravitational
waves, as predicted by general relativity (GR): $-(3.7^{+1.5}_{-1.3})
\times 10^{-14}$s/s. This period derivative is about 60 times smaller
than what was measured for the Hulse-Taylor binary pulsar \cite{wt03}.

Fortunately, the timing precision for PSR J1738+0333 is such that
we can already measure this value after four years of timing,
although not with much significance: it is $-(4.9 \pm 2.2) \times
10^{-14}$s/s. What is more important, the {\em difference} between the
predicted value and the observed value (after acorrection for
kinematic effects) is very small: $\bigtriangleup
(\dot{P_b}) = (1.7 \pm 2.7) \times 10^{-14} (< 4.4  \times 10^{-14})$.

This is the tightest limit ever on a possible contribution to the
orbital decay by the emission of {\em dipolar} gravitational waves
predicted by alternative theories of gravitation. As an example, in
Brans-Dicke gravity, the emission of dipolar gravitational waves is
given by:

\begin{equation}
\label{eq:dipolar}
\left( \frac{\dot{P_b}}{P_b}\right)_D = - \frac{2}{2 +\omega_{BD}}(s_1
- s_2)^2 \left( \frac{2 \pi}{P_b}\right)^2 \frac{m_2 R}{R + 1} T_{\odot},
\end{equation}
where $T_{\odot}$ is the solar mass in time units, and $\omega_{BD}$ is
the Brans-Dicke constant; for GR this is infinite. The variable $s_n$ is the
fractional change of the gravitational binding energy (mass) of object
$n$ with a variation of the gravitational constant $G$
($s_n\,=\,\partial \ln m_n / \partial \ln G)_{N}$ at a constant total
number of baryons $N$ (see e.g. \cite{arz03}). For neutron stars,
$s_n$ depends on the equation of state, but generally it is of the
order of 0.2. In double neutron star systems, we have $s_1 \simeq
s_2$, and therefore $(s_1 - s_2)^2 \simeq 0$. This means that
$\left( \frac{\dot{P_b}}{P_b}\right)_D$ might be zero even if
$\omega_{BD}$ is finite. In the case of MSP-WD binaries like
PSR~J1738+0333, the binding energy of the WD is many orders of
magnitude smaller than the binding energy of the MSP,
therefore $s_2\,\simeq\,0$ and $(s_1 - s_2)^2\,=\,s_1^2\,\neq 0$; for
this reason they are called ``asymmetric'' binaries. This
means that if $\left(\frac{\dot{P_b}}{P_b}\right)_D$ is very small (or
zero), then $\omega_{BD}$ must very large (or infinite).

Using $\bigtriangleup (\dot{P_b})/P_b$ 
as an experimental upper limit on $\left|
\left(\frac{\dot{P_b}}{P_b}\right)_D \right|$, we obtain
$\omega > \sim 1300 (s_1/0.2)^2$ (85\% C.L.). This is very similar
to the limits derived
from Arecibo timing of PSR~J0751+1807 \cite{nss+05}. This is
not as good as the result from the Cassini spacecraft ($\omega >
40,000$, \cite{bit03}), but it is obtained in the strong-field regime,
the only that can constrain all alternative theories of gravitation.

The main result of this study if that
there is great {\em potential} for further improvement of this test.
Over the next 5(10) years, the precision in the measurement of $\dot{P_b}$
will increase by a factor of 10(40). If the component masses are
determined from the optical studies to a precision of 10\%, or better,
then the prediction of $\dot{P_b}$ will be accurate to $6 \times
10^{-15}$ s/s or better. This will be the limiting factor in the
precision of this test. If the measured value conforms to the
prediction, that will be equivalent to $ \omega > 15,000 (s_1/0.2)^2$,
an order of magnitude improvement on all previous pulsar tests.

One of the advantages of the high timing precision of PSR~J1738+0333
has been a precise measurement of the proper motion (7.106$\pm$0.013
mas/yr in RA and 4.83$\pm$0.04 mas/yr in Dec) and the parallax
(1.08$\pm$0.07 mas). This allows a very precise correction of the
kinetic effects on the orbital period derivative.

Improving the mass ratio (definitely possible by
averaging more radial velocity measurements) and using a precise
measurement of the
orbital decay will be used to determine the mass of the pulsar and the
companion very accurately, assuming that general relativity
applies. This might be also be used to help calibrate the optical
methods for determining WD masses from their spectrum.
If it is high, the pulsar mass might be important for the study of the equation
of state for dense matter. PSR J1738+0333 might therefore be a great
physics laboratory, relevant both for the study of gravitation and the
study of the equation of state.


\begin{theacknowledgments}
The Parkes Radio Telescope is part of the Australia Telescope,
which is funded by the Commonwealth of Australia for operation
as a National Facility managed by CSIRO. The Arecibo Observatory,
a facility of the National Astronomy and Ionosphere
Center, is operated by Cornell University under a cooperative
agreement with the National Science Foundation. This research was
performed while BAJ held a National Research Council Research
Associateship Award at the Naval Research Laboratory (NRL). Basic
research in radio astronomy at NRL is supported by the Office of Naval
Research.
\end{theacknowledgments}



%
%

\end{document}